\documentclass{article}  
\usepackage{breckenr2005}
\frompage{000} \topage{000}                                              
\def\rightmark{Proposal for a High Energy Nuclear Database}
\def\leftmark{D.A. Brown et al.}

\newcommand{\url}[1]{{\tt #1}}
 
\title{Proposal for a High Energy Nuclear Database} 
\authors{
    {David A. Brown$^1$ and Ramona Vogt$^{2}$}\\[2.812mm]
    {
        \normalsize
        \hspace*{-8pt}$^1$ Lawrence Livermore National Laboratory,  
        Livermore, CA, USA\\[0.2ex] 
        \hspace*{-8pt}$^2$ Lawrence Berkeley National Laboratory, 
        Berkeley, CA, USA \\ and \\
	Department of Physics, UC Davis, Davis, CA, USA
    }
}
 
\abstract{We propose to develop a high-energy heavy-ion experimental database and make it accessible to the scientific community through an on-line interface. This database will be searchable and cross-indexed with relevant publications, including published detector descriptions.  Since this database will be a community resource, it requires the high-energy nuclear physics community's financial and manpower support.  This database should eventually contain all published data from Bevalac, AGS and SPS to RHIC and LHC energies, proton-proton to nucleus-nucleus collisions as well as other relevant systems, and all measured observables.  Such a database would have tremendous scientific payoff as it makes systematic studies easier and allows simpler benchmarking of theoretical models to a broad range of old and new experiments.  Furthermore, there is a growing need for compilations of high-energy nuclear data for applications including stockpile stewardship, technology development for inertial confinement fusion and target and source development for upcoming facilities such as the Next Linear Collider.  To enhance the utility of this database, we propose periodically performing evaluations of the data and summarizing the results in topical reviews.}

\keyword{RHIC, heavy-ion physics, database} 
\PACS{89.20.Ff, 89.20.Hh, 25.75.-q}
 
\begin{document}
 
\setcounter{page}{1}
\maketitle

\section{Background and Potential Impact}

We propose to create and maintain a high-energy nuclear database.  This central database will be web-accessible and searchable.   As with Evaluated Nuclear Data File (ENDF/B) and EXchange FORmat (EXFOR) databases \cite{bib1} and HEPDATA website \cite{bib2}, we will store cross sections, particle yields, and single particle spectra.  We will also store data specific to higher energy reactions such as multi-particle spectra, flow and correlation observables.  In other words, we will seek to archive whatever is needed to characterize a high-energy heavy-ion reaction.  Initially we will focus on published measurements but eventually we will cross-link the data with experiment descriptions.  We also envision evaluating high-energy nuclear data and reporting these results in  periodic topical reviews of subsets of the data.  The idea of publishing the topical reviews has already sparked the interest of a few review journals.  

The utility of such a database is clear: it would organize existing data, allowing easier cross-experiment comparisons, theory benchmarking and development of systematics.  In addition to the basic science needs, there is a growing list of applications for high energy nuclear data.  We show a short list of potential applications in Table 1.  Most applications do not use the data directly.  Instead, evaluated representations of the data are accessed by application codes.  

\begin{table}[htb] 
\vspace*{-12pt}
\caption[]{Some applications of high energy nuclear data.}\label{tab1}
\vspace*{-14pt}
\begin{center}
\begin{tabular}{ll}
\hline\hline\\[-10pt]
Application               & Relation to heavy ion data \\ 
\hline\hline\\[-10pt]
Proton radiography        & $pA$ data are needed to understand \\
                          & backgrounds caused by particle \\
                          & production in proton radiographs.\\\hline
Heavy-ion driven inertial & $p$ or $A$ beams on gold hohlraums drive\\
confinement fusion        & compression and eventual fusion reactions.\\\hline
$\nu$ and $\mu$ source development 
                          & $pA$ spallation reactions create particles \\
                          & for secondary beams for use in the Main Injector \\
                          & Neutrino Oscillation (MINOS) experiment and \\
                          & the Next Linear Collider (NLC).  \\\hline
Cosmic ray dose rates     & Nuclei form a large component of cosmic \\
                          & radiation.  NASA seeks to understand effects of \\
                          & dosages on humans and equipment during long \\ 
                          & term space missions.\\
\hline\hline 
\end{tabular}
\end{center}
\end{table}

Surprisingly, there is no national or international effort to collect and maintain such a database.  The US Nuclear Data Program (USNDP) \cite{bib1} has compiled low-energy nuclear reaction data for decades in the ENDF/B and other databases.  Similarly, the high-energy particle physics community is served by the HEPDATA, PDG, arXiv.org and SLAC-SPIRES websites.  Table 2 summarizes these databases.  The high-energy nuclear physics community is only partially served by these data sources.  One could argue that most experiments make their published data available - for example, PHENIX posts tables of published data on the collaboration website.  Inevitably this leads to a proliferation of data formats and web sites.  Furthermore, experiments end and their web servers may no longer be maintained.  Thus, there is a very real risk that the data could be lost.
Given the volumes of data generated by experiments at the RHIC and future experiments at the LHC, GSI and elsewhere, this oversight should be rectified.

\begin{table}[htb] 
\vspace*{-12pt}
\caption[]{Related database efforts.  Derived databases (such as NuDat) are not included in this table.}\label{tab2}
\vspace*{-14pt}
\begin{center}
\begin{tabular}{ll}
\hline\hline\\[-10pt]
Database &  Description\\ 
\hline\hline\\[-10pt]
ENDF/B \cite{bib1}            & Evaluated nuclear reaction data \\\hline
EXFOR \cite{bib1}             & Experimental nuclear reaction data \\\hline
Evaluated Nuclear Structure   & Nuclear structure information \\
Data File (ENSDF) \cite{bib1} & \\\hline
Reference Input Parameter     & Nuclear reaction modeling parameters \\
Library (RIPL) \cite{bib1}    & \\\hline
Computer Index Neutron        & Nuclear physics publication database\\
DAta (CINDA) \cite{bib1}      & \\\hline
Nuclear Science               & Nuclear physics publication database\\
References (NSR)\cite{bib1}   & \\\hline
HEPDATA \cite{bib2}           & High-energy physics (HEP) reaction data \\\hline
Particle Data Group (PDG) \cite{bib3} & Particle properties \\\hline
SLAC-SPIRES \cite{bib4}       & HEP publication database \\\hline
arXiv.org \cite{bib4.5}        & Preprint database \\
\hline\hline 
\end{tabular}
\end{center}
\end{table}

In the next few sections, we explain that the proposed database should be a community effort, motivate the need for data evaluation and topical reviews and provide some technical details.  We conclude with a status of the database proposal. 

\section{Database Management Philosophy}

Since this database would be a community resource, we propose a community driven management model such as the arXiv.org preprint server -- the ``consumers'' of arXiv.org are also its ``suppliers.''   On one hand, physicists submit their preprints to arXiv.org because it serves as a form of advertising.  On the other hand, they browse arXiv.org because they know others are submitting their latest results there.  In this way, data collection is farmed out to the data producers -- a tactic we wish to employ.  

The proposed database would differ from arXiv.org in two key respects.  First, the proposed database would not only contain published data but also auxiliary or supporting data sets that may be too large for publications such as Physical Review Letters.  Second, in order to assure that we only have high-quality data, we would like to piggyback on various journals' peer-review processes.  Ideally this means that authors will submit both the published and auxiliary supporting data to the database when submitting papers for publication.  One submission model would be to add links directly to journal submission pages.  Preliminary discussions with journal editors indicate the willingness of the journals to cooperate in this endeavor.

For the experimental collaborations to have the political will to support this project, they must be given both a financial stake in the eventual product and have a hand in steering the database development.  To encourage this, we propose holding annual workshops to guide development as well as discuss the topical reviews and propose new subjects for review. 

\section{Evaluations and Topical Reviews}\label{review}

Evaluated data is our ``best guess'' representation of a particular observable/re-action/etc.  Whether one obtains this evaluation through a model calculation, systematics or a fit to raw experimental data, the final product needs to be checked against existing data and it needs to be peer reviewed.  However the evaluation is produced, we need a better understanding of the raw experimental data.  We illustrate this with two case studies: $D$ meson production data and two-particle correlation data.  

\subsection{Case Study 1: D Meson Production}

\begin{figure}[htb]
\insertplot{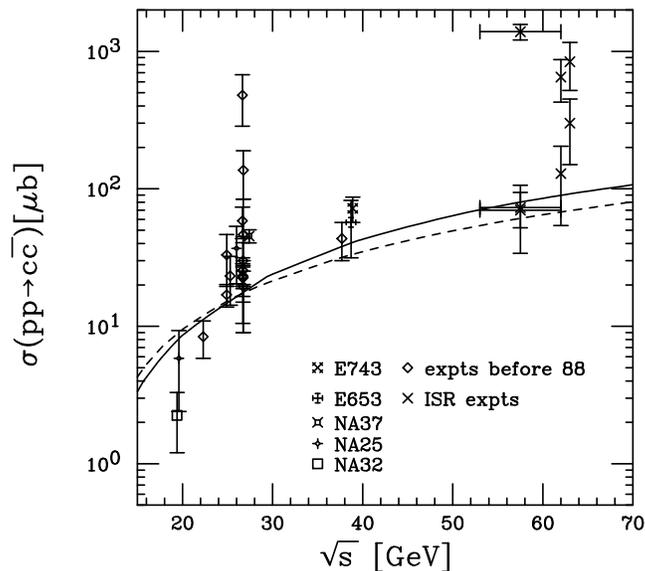}
\vspace*{-1cm}
\caption[]{Total charm production cross sections from $pp$ and $pA$ measurements compared to calculations with the MRST HO parton densities using $m_c$  = 1.2 GeV and $\mu$ = 2$m_c$ (solid) and GRV98 HO parton densities using $m_c$ = 1.3 GeV and $\mu$ = $m_c$ (dashed). Updated from Ref. \cite{bib5}.}
\label{fig1}
\end{figure}

The $D$ meson, a bound state of a charm quark and a light antiquark, is a hot new topic in heavy ion physics.  Because charm production is calculable in perturbative quantum chromodynamics (pQCD), it is possible to predict medium effects such as shadowing \cite{bib6}, energy loss \cite{bib7} and transverse flow \cite{bib8} on the $D$ meson.  All of these predictions are relevant for the understanding of the quark-gluon plasma. $D$ mesons have been measured in a large number of experiments over a wide energy range in $pp$ and $pA$ interactions.  To predict $D$ meson yields within the parton model, one must understand the total charm production cross section and the appropriate fragmentation functions.  As seen in Fig.~1 from Ref.~\cite{bib5}, the measured total charm cross section data varies considerably, even at a specific energy.  The solid and dashed curves are examples of recent theoretical calculations at the limit of applicability of pQCD.  It is common for users to throw out a significant fraction of this data according to their own prejudices.  At some level this is understandable since the early measurements are often too high due to poor statistics and unreasonable extrapolations to full phase space.  The early RHIC results are also above the extrapolation of these curves to higher energy \cite{bib9}.  This discrepancy needs to be better understood, both from the point of view of the latest results and the previous measurements.  Therefore, the previous data must be re-evaluated, accounting for the number of events measured, the acceptance, assumptions due to phase space extrapolation, the $D$ decay branching ratios used and any corrections for $A$ dependence.   There is a clear need for both a central database and for data evaluators.  

\subsection{Case Study 2: Two-particle Correlations}

\begin{figure}[htb]
\epsfclipon\insertplot{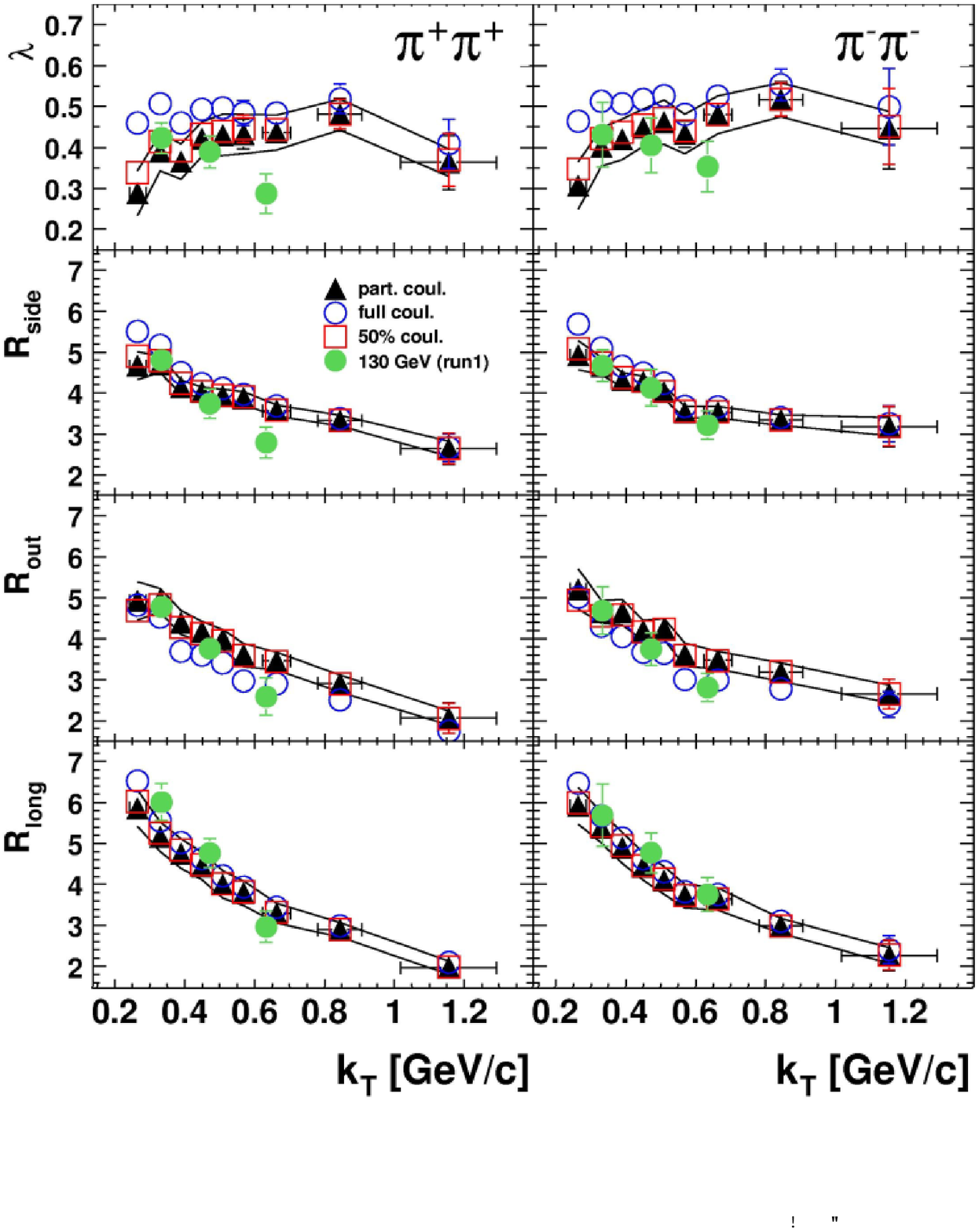}
\vspace*{-1cm}
\caption[]{HBT radii and intercepts, $\lambda$, from Gaussian fits to two pion correlations in 200 GeV Au+Au collisions as measured by the PHENIX collaboration.  The filled circles are data from the 130 GeV run in RHIC year 1.  The other three data sets are 200 GeV data using three separate Coulomb correction schemes.}
\label{fig2}
\end{figure}

One of the most surprising results from the first few years of RHIC running is the so-called ``HBT Anomaly.''  Hanbury Brown/Twiss (HBT) interferometry is an experimental technique that uses the interference patterns in two-particle spectra caused by quantum statistics to access to the space-time development of heavy-ion collisions.   By carefully fitting the two-particle correlation function (the ratio of the true pair spectrum to an uncorrelated background spectrum), one can extract the ``HBT radii.'' These radii are a rough measure of the space-time extent of the system when the particles of interest were emitted.  The RHIC ``HBT Anomaly'' has two aspects.  First, the pion HBT radii measured at RHIC seem to indicate a zero or nearly zero freeze-out duration, $\Delta\tau$.  In many models, $\Delta\tau$ is controlled by the HBT radii $R_{\rm out}$ and $R_{\rm side}$.  As we see in Fig.~2, PHENIX fits from Ref. \cite{bib10} indicate that $R_{\rm out}\approx R_{\rm side}$, suggesting that $\Delta\tau \approx $ 0 fm/$c$ over the whole range of pair transverse momentum.  The second part of the ``HBT Anomaly'' is the extremely weak evolution of the pion HBT radii over two decades of bombarding energy.   Both conclusions hinge on fits of the HBT radii, yet it is known that the radii are very sensitive to Coulomb corrections and other experimental correction procedures.  Indeed, the triangles, open circles and open squares in Fig.~2 indicate the sizable variation in $R_{\rm out}$, $R_{\rm side}$, $R_{\rm long}$ and $\lambda$ due to the various Coulomb corrections.  Ideally one would like all of the true-pair and mixed-pair background spectra from all published two-particle correlation data sets, not just same-sign pion data, collected in one place so that others can perform their own analyses and draw their own conclusions.  Furthermore, this collection should not be limited to heavy-ion data since various statistical models suggest a deep connection between the two-particle sources at RHIC and those in $e^+ e^-$ and $p \bar{p}$ collisions.  

\section{Technical Details}

Many of the tools that we would need to produce this database are available ``off-the-shelf.''  We envision that users will access the database through a set of JSP or PHP dynamic web pages.  Both JSP and PHP have tools that simplify on-line database queries.  The data itself will be a MySQL relation database.  We also envision that the dynamic web pages will be able to send the data directly to a plotting utility such as LLNL's LLNLPlot Java applet.  LLNLPlot can plot 2D and 3D data and is the plotting back-end of the Nuclear and Atomic Data System (NADS) \cite{bib11}.

A central, yet often neglected, aspect of data archives is the technical details of the data storage format.  The nuclear data community traditionally suffers from a multitude of relatively obscure data formats. For example, the format still used in the ENDF/B database was designed specifically to accommodate the limitations of now-obsolete punch cards.  In some cases, the task of writing translation and visualization tools for these data sets requires a large, dedicated effort.

Given the importance of using a transparent and well-supported format, we have decided on XML (eXtensible Markup Language) as our data storage format. Documents stored in XML can be self-describing so that, with minimal effort, scientists/users 30 years from now can interpret the documents' contents.  Furthermore, XML documents are represented by computationally convenient tree structures rather than the simple strings typically used to store nuclear data.  XML is a mature technology with the support of thousands of programmers and web developers and is extensively supported by most common programming languages.  Lastly, the many tools needed for web-based access and manipulation of XML databases have reached a state of maturity.

\section{Current Status}

We have submitted a white paper describing our proposal to the DOE-OS Heavy-Ion and Nuclear Theory programs and are circulating it in the STAR and PHENIX collaborations.  Copies of the whitepaper are available upon request from the authors.  Due to the budget cuts in FY05-06, it is unlikely to be funded before FY07.  Despite the funding uncertainties plaguing the field, elements of this proposed project are in the process of being developed for other uses, namely the XML data format and the LLNLPlot plotting tool.  In Fig. 3, we present a timeline for the remaining parts of the proposal.  Since we want this database to be a community resource, we strongly encourage members of the heavy-ion community to contact us with their questions, comments, wishes and ideas.  

\begin{figure}[htb]
\insertplot{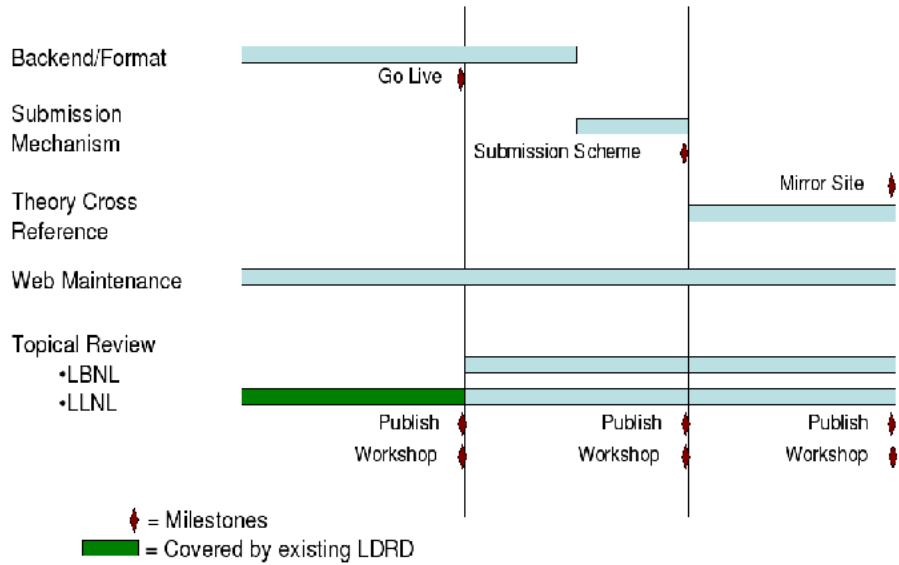}
\vspace*{-1cm}
\caption[]{Project timetable.}
\label{fig3}
\end{figure}

\section*{Acknowledgment(s)}

This work was performed under the auspices of the U.S. Department of Energy by University of California, Lawrence Livermore National Laboratory under Contract W-7405-Eng-48. R.V. was supported in part by the Director,
Office of Energy Research,
Office of High Energy and Nuclear Physics,
Nuclear Physics Division of the U.S.\ Department of Energy
under Contract No.\ DE-AC03-76SF00098.

\vfill\eject
\end{document}